# Room-temperature vacuum Rabi splitting with active control in two-dimensional atomic crystals

Jinxiu Wen,[1,3†] Hao Wang,[1,3†] Weiliang Wang,[1,3] Zexiang Deng,[3] Chao Zhuang,[1,2] Yu Zhang,[1,2] Fei Liu,[1,2] Juncong She,[1,2] Jun Chen,[1,2] Huanjun Chen,[1,2*] Shaozhi Deng[1,2*] & Ningsheng Xu[1,2*]

[1]State Key Laboratory of Optoelectronic Materials and Technologies, Guangdong Province Key Laboratory of Display Material and Technology, Sun Yat-sen University, Guangzhou 510275, China.

[2]School of Electronics and Information Technology, Sun Yat-sen University, Guangzhou 510006, China.

[3]School of Physics, Sun Yat-sen University, Guangzhou 510275, China.

*e-mail: chenhj8@mail.sysu.edu.cn; stsdsz@mail.sysu.edu.cn; stsxns@mail.sysu.edu.cn.

†These authors contributed equally to the work.

**Abstract**: Strong light–matter coupling manifested by vacuum Rabi splitting has attracted tremendous attention due to its fundamental importance in cavity quantum-electrodynamics research and great potentials in quantum information applications. A prerequisite for practical applications of the strong coupling in future quantum information processing and coherent manipulation is an all-solid-state system exhibiting room-temperature vacuum Rabi splitting with active control. Here we realized such a system in heterostructure consisting of monolayer $WS_2$ and an individual plasmonic gold nanorod. By taking advantage of both of the small mode volume of the nanorod and large binding energy of the $WS_2$ exciton, giant vacuum Rabi splitting energies of 91 meV ~ 133 meV can be obtained at ambient conditions, which only involve 5 ~ 18 excitons. The vacuum Rabi splitting can be dynamically tuned either by electrostatic gating or temperature scanning. These findings could pave the way towards active quantum optic devices operating at room temperature.

Strong coupling occurs when the energy exchange between a quantum emitter and optical cavity is fast enough to overcome their individual dissipation rates. In such a scenario, mixed states with part-light and part-matter characteristics will be formed and vacuum Rabi splitting (VRS) can be observed in the optical spectra of the hybrid[1,2]. In this regime, the fast and coherent energy exchange between the emitter and cavity allows for test-bed study of several fundamental problems in quantum-cavity electrodynamics, such as quantum entanglement, decoherence processes, and Bose–Einstein condensate[3–6]. From the application perspective, the strong coupling effect also opens up new avenues for quantum manipulation, quantum information storage and processing, as well as ultrafast single-photon switches[7–10]. For these applications a robust, all-solid-state system exhibiting room-temperature, large VRS with active control is strongly desired. Furthermore, strong coupling involving only a few and even a single emitter is a pivotal requirement to access the quantum optics effects[6]. So far several systems have been demonstrated with strong coupling, including the atoms in optical cavity[11], quantum dots in microcavity and photonic crystals[2,12], quantum well sandwiched between distributed Bragg reflectors[13], and plasmonic nanostructures coupled with molecules or quantum dots[14–16]. In comparison with the other counterparts, the plasmonic nanocavities can remarkably facilitate the strong coupling process in terms of their deep subwavelength mode volumes, simple excitation manner, and facilely tunable resonance frequencies. These merits allow for strong coupling at ambient conditions with large VRS upto ~ 1 eV[17]. Despite these progresses, it is still difficult to access the active control of the strong coupling in these systems because the tunability of the plasmonic nanocavities and excitons are rather limited. On the other hand, to push the interactions into single-exciton level, delicately designed architectures have to be implemented into these systems[15, 18], whereby the robustness of the strong coupling is severely compromised.

Two-dimensional transition-metal dichalcogenides (TMDs) have attracted much attention in recent years due to the exotic optical and optoelectronic responses associated with their atomically thin thicknesses[19]. In comparison with the molecules or quantum dots, the TMDs possess several features which can greatly favor the strong coupling. First, the monolayer TMDs are usually direct-bandgap semiconductors with large exciton binding energies (0.3 eV ~ 0.9 eV) and exciton transition strengths[20]. Second, the exciton characteristics of the TMDs can be facilely manipulated by external stimulus, such as electrostatic gating, thermal scanning, and optical pumping[19, 21–23]. Third, the single-crystalline TMDs exhibit uniform electronic and optical properties across the entire two-dimensional flake. Such uniformity can greatly facilitate and guarantee the robustness of the strong coupling effect. Last but not the least, the intrinsic valley-controlled optical properties in the TMDs are foreseen to endow the strong coupling with valley functionalities, which can open up new avenues for both of fundamental research and practical applications. Although a few studies have demonstrated the potential of the TMDs for strong coupling[24–28], they all involved plenty of excitons. Furthermore, active control on the strong coupling by taking advantage of the exceptional excitonic properties of the TMDs has yet to be demonstrated.

Here, to the best of our knowledge, we for the first time realize all-solid-state, room-temperature strong coupling with active control in heterostructure consisting of monolayer $WS_2$ and an individual plasmonic gold nanorod. The nanorod can confine the electromagnetic field into an ultrasmall volume, whereby the coherent interaction between the two-dimensional $WS_2$ exciton and plasmon resonance can give rise to strong coupling with VRS of upto ~ 100 meV at room temperature. The number of excitons involved is about 5 ~ 18, which approaches the quantum optics limit. In addition, the strong coupling can be dynamically tuned either by electrostatic gating or temperature scanning.

**Results**

**Realization of room-temperature strong coupling.** The $WS_2$ sample was grown by chemical vapor deposition (CVD) method, which is single-crystalline flake with triangular shape (Fig. 1a and Supplementary Fig. 1a). The monolayer nature of the $WS_2$ can be confirmed by both of atomic force microscopy (AFM) and Raman characterizations (Fig. 1b and Supplementary Fig. 1b), whereby the thickness of a typical flake can be determined as ~ 1 nm. The pristine monolayer $WS_2$ exhibits an exciton emission centering at 1.950 eV, with a narrow line width ($\gamma_{ex}$) of 57 meV (Fig. 1c, upper panel). The photoluminescence and Raman intensity mappings across the entire flake reveal excellent uniformity of the sample (Fig. 1d and Supplementary Fig. 1c). Gold nanorods were chosen as plasmonic nanocavity (Fig. 1e), which exhibit two plasmon resonance modes associated respectively with electron oscillations along the transverse (TPM) and longitudinal (LPM) directions of the nanorods. We focused on the LPM, which can be synthetically tailored by tuning the aspect ratios of the gold nanorods[29]. On the other hand, we employed the single-particle dark-field scattering technique to rule out the average effect from the ensemble measurements as well as to realize few-exciton strong coupling from an individual gold nanorod.

To achieve the strong coupling the gold nanorods were sparsely deposited onto the monolayer $WS_2$ flake by drop-casting (Fig. 1f). As shown in Fig. 1c (middle panel), a typical gold nanorod with aspect ratio of 2.2 exhibits a well-defined LPM centering at 1.970 eV, which is in resonance with the exciton emission of the $WS_2$. In such a scenario, mode splitting can be clearly observed on the scattering spectrum from the gold nanorod coupled to the $WS_2$, where the high- (HEM) and low-energy (LEM) hybrid modes are formed (Fig. 1c, lower panel). To ascertain that the gold nanorod−$WS_2$ heterostructure enters the strong coupling regime, we measured the scattering spectra from various individual nanostructures with different detunings between the plasmon frequencies and exciton transition. To that end, gold nanorods with different LPM energies were prepared by tailoring their aspect ratios (Supplementary Fig. 2a−c). Subsequently they were coupled to the same $WS_2$ flake for measuring the scattering spectra. As the LPM energies were progressively varied across the exciton transition energy, the two prominent hybrid modes always existed on the scattering spectra of the heterostructures. In particular, when the LPM frequencies were detuned from the high-energy side of the exciton transition, the HEM dominates and redshifts with decreasing LPM frequencies. As the LPM becomes overlapped with the exciton transition, the intensities of the HEM and LEM are comparable with each other. When the LPM moves away from the exciton transition to the low-energy side, the LEM overwhelms the HEM and redshifts as the detunings become larger (Fig. 2a and Supplementary Fig. 2d). These experimental spectra can be further corroborated using electromagnetic simulations, where the geometries of the heterostructures were set according to their SEM images (Supplementary Fig. 3). As shown in Fig. 2b, the calculated spectral shapes and evolvements agree well with the experimental ones. The small discrepancy on the scattering dip position is due to the dielectric function of the $WS_2$ used, which was measured on samples obtained by mechanical exfoliation. The exciton transition energies can be violated between the CVD-grown and exfoliated $WS_2$ flake due to the presence of impurities or defects. Such discrepancy will not disturb the underlying physics of the mode splitting.

The spectral evolution of the heterostructures with the energy detunings can be seen more clearly on the normalized scattering spectrum diagram (Fig. 2c). As the detunings are varied across the zero point, the scattering spectra show a distinct anti-crossing behavior with two prominent branches associated respectively with the

HEM and LEM. The anti-crossing behavior is a typical characteristic of the strong coupling between the plasmonic cavity and exciton transition[6]. A coupled harmonic oscillator model was then utilized to describe the strong coupling effect, which can reproduce the experimental data very well. The Rabi splitting energy, $\hbar\Omega$, can be readily extracted to be 106 meV at zero detuning. To the best of our knowledge, this is the largest VRS observed at room-temperature by far in TMDs[24–28]. The plasmon line width of the gold nanorods can be extracted as ~ 149 meV according to the algorithm proposed previously[14]. As a result, we find that the Rabi splitting energy fulfill the criterion where the strong coupling can occur ($\hbar\Omega > \frac{\gamma_{pl} + \gamma_{ex}}{2}$)[24]. It should be noted that the scattering spectra with two split modes separated by the dip at the exciton transition may arise from exciton-induced transparency or enhanced absorption rather than the strong coupling[30]. To address this argument, more calculations were performed to obtain the absorption and extinction spectra of the heterostructures corresponding to the Fig. 2a and b. The spectral dips as well as anti-crossing behavior always exist on these two types of spectra (Supplementary Fig. 4), indicating a true strong coupling to occur. In the strong coupling regime, hybridized modes with part-light and part-matter will be formed due to the coherent energy exchange between the exciton and plasmon resonance. We then utilized the coupled harmonic oscillator model to calculate the respective contributions from exciton and plasmon components. As shown in Fig.2d, by detuning the plasmon resonance to low-energy side of the exciton transition, the LEM is more plasmon-like while the HEM is exciton-like, and *vice versa* for detuning to the high-energy side. The strong coupling is also rigorously dependent on the separations between the nanorod and WS$_2$. As the gold nanorod is kept away from the WS$_2$, the split modes will vanish quickly and disappear for separations larger than 7 nm (Supplementary Fig. 5). Such a behavior suggests that the coupling is mediated by the plasmonic near-field located within the gap between the nanorod and WS$_2$.

An important consequence of the strong coupling is that the exciton should carry the plasmonic characteristics *via* the coherent interactions. This can be revealed by polarization spectroscopy. The scattered light of the LPM is linearly polarized along the length direction of the nanorod (Supplementary Fig. 6a). The light scattered by the pristine WS$_2$ exciton is isotropically polarized (Fig. 2e), which is due to the spherical symmetry of the 1s-state exciton wavefunction[31]. In contrast, at zero detuning the two split modes with equivalent magnitudes followed the same polarization states of the LPM. Their polarization states also varied with different detunings. When the LPM energy is reduced away from the exciton, the polarization degree of the LEM is almost undisturbed, whereas that of the HEM is evidently suppressed (Supplementary Fig. 6b). If the detunings is varied to the high-energy side of the exciton, the polarization of the HEM is invariant while the LEM next to the exciton transition becomes less polarized (Supplementary Fig. 6c). By defining the degree of polarization (DOP) as $\rho = (I_{max}-I_{min})/(I_{max}+I_{min})$, where $I_{max}$ and $I_{min}$ are the maximum and minimum light intensities at two perpendicular polarization directions, we can quantitatively analyze the scattering polarizations. As the detunings are varied across the zero point from the minus direction, the ratios of the DOP between the LEM and

HEM is reduced gradually (Fig. 2f, purple spheres), suggesting that the LEM evolves from plasmon-like into exciton-like resonance. This evolvement can be further corroborated by plotting the ratios between the plasmon fractions of the LEM and HEM against the energy detunings, which exhibit similar trend as that of the DOP (Fig. 2f, blue line).

**Strong coupling approaching the quantum optics limit in the two-dimensional WS$_2$.** The facile tuning of the nanorod geometries allows for studying the few-emitter strong coupling, which is very important for quantum information applications. To establish such interactions one needs to create nanocavities with ultrasmall mode volumes where several and even only one exciton can occupy. We realize such nanocavities by precisely tuning the nanorod volume while fixing the plasmon resonance frequency at the exciton energy. As evidenced in Fig. 3a, when the volume of the nanorod is reduced, the mode splitting always exists while the $\hbar\Omega$ is weakened from ~ 133 meV to ~ 91 meV. We then estimate the number of the excitons taking part in the strong coupling by referring to the near-field distributions of the nanorods and size of the exciton. The WS$_2$ exciton is of two-dimensional with a reduced out-of-plane dipole moment[31]. We therefore calculate the "mode area", $S$, which characterizes the spreading of the plasmonic near-field on the WS$_2$ surface. As shown in Fig. 3b, the plasmonic field is strongly confined and enhanced at the WS$_2$ surface, with the strongest enhancements locating near the two apexes of the nanorod. The $S$ becomes larger when the volumes of the nanorod are increased (Supplementary Table 1). The size of the exciton can be evaluated according to its wavefunction

(BerkeleyGW package)[31, 32], which extends several unit cells (Fig. 3c). The effective area on the WS$_2$ surface occupied by the exciton, $A$, can thereafter be calculated to be 26.8 nm$^2$. Subsequently the number of excitons $N$ contributing to the $\hbar\Omega$ of a specific heterostructure can be calculated *via S/A*. The effective number $N$ of excitons coupled with the smallest gold nanorod is ~ 5. More insights can be obtained by investigating the dependence of the coupling strength $g$, which is defined by $\hbar\Omega/2$, on the $N$. According to the previous theoretical predictions, the relation between the $g$ and $N$ can be written as[18],

$$g = \mu_m \sqrt{\frac{4\pi\hbar N c}{\lambda \varepsilon \varepsilon_0 V}} \quad (1)$$

where $\mu_m$ is the transition dipole moment of the exciton, $V$, $\varepsilon$, and $\lambda$ are the mode volume, dielectric function, and wavelength, respectively. Therefore there should be linear dependence of $g\sqrt{V}$ on the $\sqrt{N}$. This is clearly evidenced in the strong coupling between the plasmon and WS$_2$ exciton (Fig. 3d).

**Active control on the strong coupling in the two-dimensional WS$_2$.** We finally demonstrate the active control on the strong coupling by taking advantage of the WS$_2$ exciton that is amenable to external stimulus. One notes that the WS$_2$ exciton is sensitive to the temperature variations (Supplementary Fig. 7a, b)[33], while the plasmon resonance energy of the gold nanorods shows much weaker dependence on the environmental temperature (Supplementary Fig. 7c). The resonance coupling can therefore be stimulated by scanning the temperatures in a specific heterostructure. As the temperature was varied from 293 K to 433 K, the WS$_2$ exciton can be scanned across the plasmon resonance of the gold nanorod. In such a manner, the anti-crossing of the exciton transition with the plasmon resonance can be induced (Fig. 4a, b, and Supplementary Fig. 8). The two hybrid branches appear and repel each other as the detuning between the exciton energy and plasmon resonance is varied from negative to positive, showing that the interaction between the plasmon and exciton can be tuned between the strong coupling and weak coupling regimes. The extracted Rabi splitting energy can be up to 110 meV.

On the other hand, the tunability of the exciton states in TMDs by electrical charging has offered unprecedented potentials in active devices made with TMDs. In comparison with the thermal stimulation, electrical control is more preferred in terms of its facile implementation, high speed, and compatibility with the state-of-art integrated circuit techniques. We therefore performed the electrical control on the strong coupling in the gold nanorod−$WS_2$ heterostructures. To that end, the gold nanorods were deposited onto a field-effect transistor (FET) fabricated with the monolayer $WS_2$ (Supplementary Fig. 9). The $WS_2$ can be electrostatically doped by applying different gate voltages ($V_g$) to the FET, whereby the scattering and photoluminescence spectra from a specific heterostructure can be recorded. With positive $V_g$, the photoluminescence intensities and peaks of the exciton transition are almost invariant upon increasing voltages. In contrast, the photoluminescence becomes weaker and redshifts as the $V_g$ is increased towards the negative side (Supplementary Fig. 10). These dependences can be ascribed to formation of charged excitons, i.e. trions, and state-blocking effect[34]. The gate-dependent measurements suggest the possibility of tuning the strong coupling under negative $V_g$. Shown in Fig.

4c is the contour plot of the evolution of the scattering spectra from the heterostructure, where the $V_g$ is alternately switched between 0 and −120 V and the scattering spectra under each bias are stacked together. Upon the negative bias, the exciton transition is suppressed and the strong coupling is thereafter modulated (Supplementary Fig. 11). Specifically, the intensity ratio between the LEM and HEM is reduced and the scattering dip separating them becomes shallower. Furthermore, the LEM red-shifts ~ 5 meV. When the bias is removed, the scattering spectrum of the strong coupling recovered gradually. It is noted that the LEM is more sensitive to the gate voltage than the HEM. This is due to that the doped electrons by the negative bias will occupy the low-energy states first. As a result, the optical transition associated with these states will be suppressed more severely than that of the high-energy states. Besides, as the exciton energy is shifted to lower energy, the LEM is more exciton-like and thereby sensitive to the applied bias. To further demonstrate the active control ability, the scattering-dip intensity and peak of the LEM ($E_{P-}$) are extracted from the scattering spectra, whereby their cyclic performances are investigated upon alternating $V_g$. As shown in Fig. 4d, the intensities of the scattering-dip relative to the scattering maximum change between ~ 0.76 and ~ 0.65 with switching on and off the bias. The LEM maxima shift back and forth between ~ 1.936 eV (off) and ~ 1.928 (on). Although the modulation depths are relatively small, these observations unambiguously indicate the tunability of the strong coupling in the heterostructure by the external gating.

**Discussion**

We have succeeded in demonstrating all-solid-state, room-temperature, and active controllable strong coupling in monolayer $WS_2$ coupled to a single plasmonic gold nanorod. The obtained VRS ranges from 91 meV ~ 133 meV, where only 5 ~ 18 excitons were involved. The heterostructures proposed constitute potential candidate for future on-demand quantum optics devices.

## Methods

### Sample preparation

The gold nanorods were synthesized using the seed mediated method[29]. The LPM wavelength of the as-grown gold nanorod solution was 700 nm. The as-prepared sample was subjected to anisotropic oxidation[29]. Real-time monitoring of the oxidation process was conducted by measuring the extinction spectra of the gold nanorod solution. To prepare the heterostructure the gold nanorods were deposited onto the monolayer $WS_2$ flake by drop-casting. Various heterostructures can be obtained after the deposit was dried naturally under ambient conditions.

### Spectroscopy

The scattering spectra of the heterostructures were recorded on a dark-field optical microscope (Olympus BX51) that was integrated with a quartz-tungsten-halogen lamp (100 W), a monochromator (Acton SpectraPro 2360), and a charge-coupled device camera (Princeton Instruments Pixis 400BR_eXcelon). The camera was thermoelectrically cooled down to −70 ºC during the measurements. A dark-field objective (100×, numerical aperture 0.80) was employed for both illuminating the heterostructures with the white excitation light and collecting the scattered light. A linear polarizer (U-AN360 Olympus, JAPAN) was placed in the optical path before the monochromator for measuring the polarization-dependent scattering spectra. The polarization axis of the polarizer was aligned horizontally. The extinction spectra of the aqueous gold nanorod samples were obtained using a HITACHI U-4100 UV/visible/near-infrared spectrophotometer with an incidence spot size of 5 mm. The photoluminescence and Raman spectra of the monolayer $WS_2$ were collected using a Renishaw inVia Reflex system with a dark-field microscope (Leica). The excitation laser of 532 nm was focused onto the samples with a diameter of ~ 1 μm through a 50× objective (numerical aperture 0.8).

**Characterizations**

SEM images of the heterostructures were acquired using an FEI Quanta 450 microscope. The thickness of the monolayer $WS_2$ was measured using AFM (NTEGRA Spectra). HRTEM and SAED measurements were conducted using field emission TEM (FEI Tecnai[3] G2 60-300) with an operation voltage of 300 kV.

**Acknowledgements**

We thank Professor Wencai Ren and Mr. Yang Gao from the Institute of Metal Research at Shenyang for help in TEM sample preparations. This work was financially supported by the National Natural Science Foundation of China (Grant Nos. 51290271, 11474364), the National Key Basic Research Program of China (Grant Nos. 2013CB933601, 2013YQ12034506), the Guangdong Natural Science Funds for Distinguished Young Scholars (Grant No. 2014A030306017), and the Guangdong Special Support Program.

**Author contributions**

N.X., S.D., and H.C. conceived the study and designed the experiments. J.W., H.W., and C.Z. conducted the experimental measurements and theoretical calculations. W.W. and Z.D. performed the DFT calculations. Y.Z., F.L., J.S., and J.C. participated in the discussion of the data. J.W, H.W., N.X., S.D., and H.C. co-wrote the manuscript.
†These authors contributed equally.

**Additional information**

**Supplementary Information** accompanies this paper at

http://www.nature.com/naturecommunications

**Competing financial interests:** The authors declare no competing financial interests.

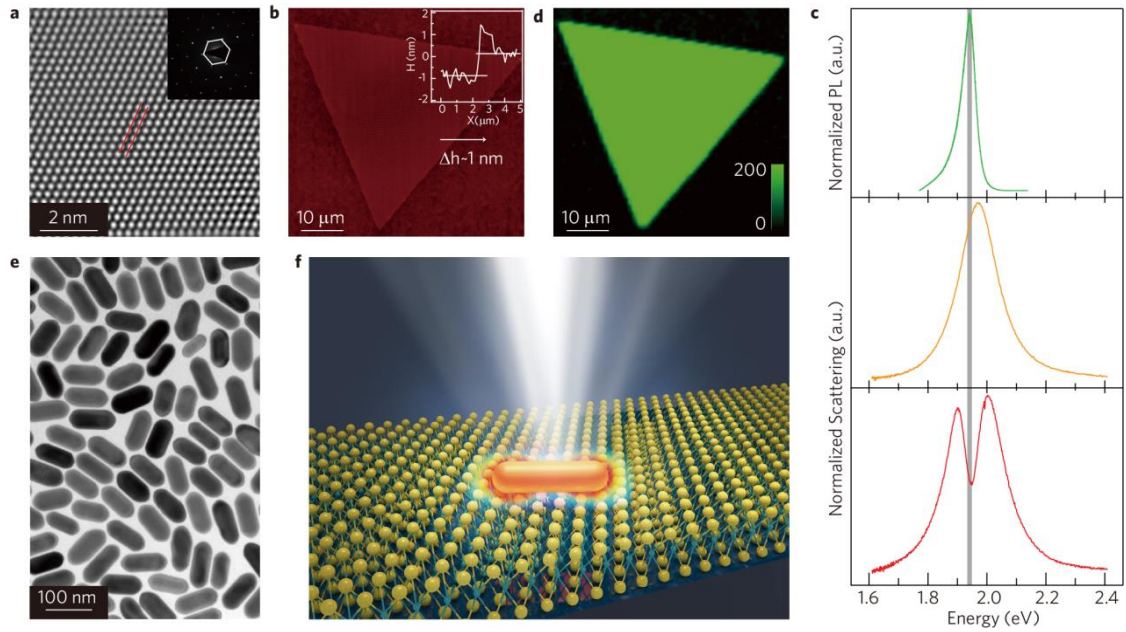

**Figure 1 | Gold nanorod−WS$_2$ heterostructures. a**, High-resolution transmission electron microscopy (TEM) image of the monolayer WS$_2$ flake. Inset showing selective area electron diffraction (SAED) pattern on the sample, which indicates a hexagonal crystal structure. **b**, AFM image of the monolayer WS$_2$. **c**, Photoluminescence spectrum of the monolayer WS$_2$ (upper), scattering spectra of a pristine individual gold nanorod (middle), and nanorod coupled to the WS$_2$ (lower). **d**, Photoluminescence mapping of the WS$_2$ flake corresponding to (**b**). **e**, TEM image of the pristine gold nanorods. **f**, Schematic showing the heterostructure composed of an individual gold nanorod coupled to the WS$_2$.

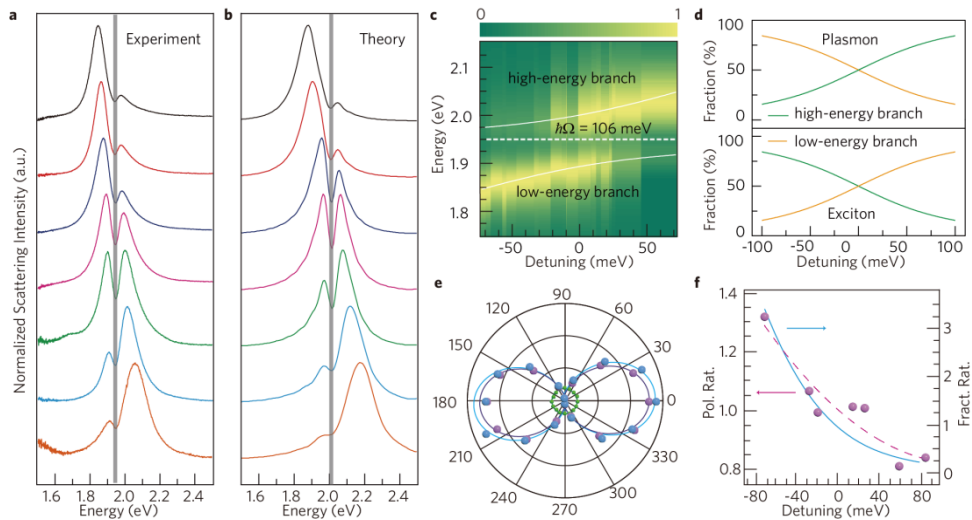

**Figure 2 | Strong coupling in gold nanorod−WS$_2$ heterostructures. a**, Dark-field scattering spectra from different individual gold nanorods coupled to the same monolayer WS$_2$ flake. The aspect ratios of the gold nanorods are 1.7 (orange), 1.9 (blue), 2.1 (green), 2.2 (pink), 2.3 (purple), 2.5 (red), and 2.6 (black), respectively. **b**, Calculated spectra corresponding to the experimental spectra shown in (**a**). **c**, Colored coded normalized scattering spectra from the heterostructures with different detunings between the plasmon resonances and exciton. The two white lines represent the fittings using the coupled harmonic oscillator model. The dashed white line indicates the exciton transition energy. **d**, Plasmonic (upper) and exciton (lower) fractions for the HEM and LEM of the heterostructures, respectively. **e**, Polarization polar plots of scattering intensities of the pristine WS$_2$ exciton (green), HEM (blue), and LEM (purple) of the heterostructure. The respective solid curves are fitting results using the cosine squared functions. **f**, Solid blue line: dependence of the ratios between the plasmonic fractions of the LEM and HEM on the detunings. Purple spheres: ratios of degree of polarization (DOP) between the LEM and HEM as a function of the detunings. The purple dashed line is guide for the eye.

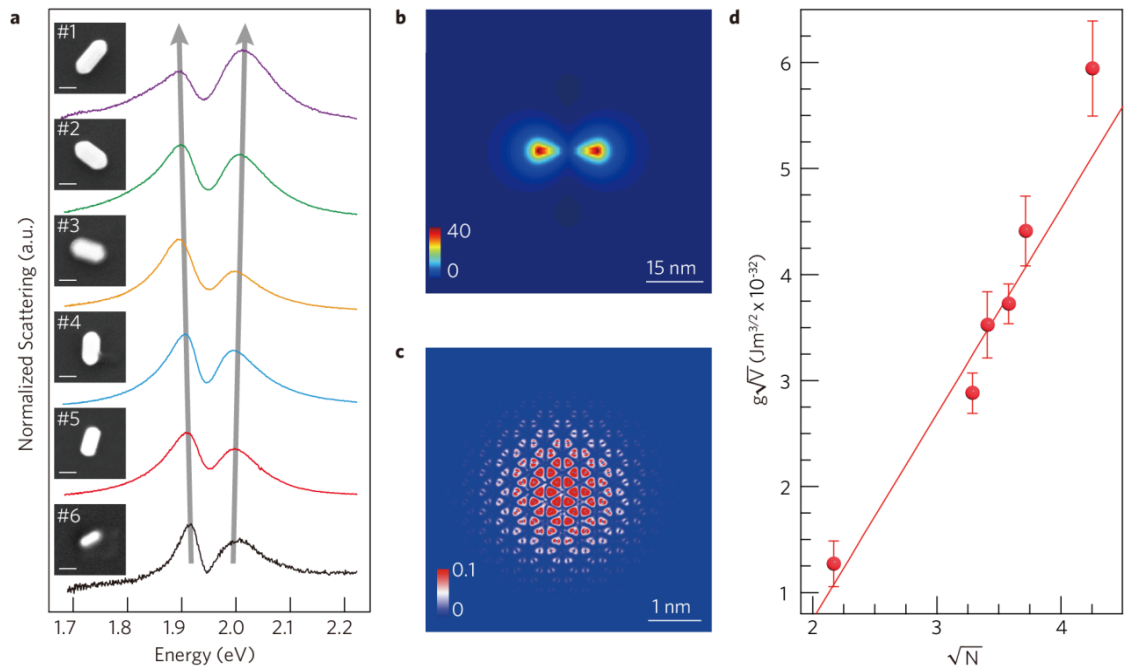

**Figure 3 | Strong coupling approaching the quantum limit. a**, Dark-field scattering spectra from gold nanorods of different volumes coupled to the same monolayer $WS_2$ flake. The scale bars are 50 nm. **b**, Near-field enhancement contour of an individual gold nanorod placed onto the $WS_2$ monolayer. The electric field contour is drawn on the $WS_2$ plane underneath the gold nanorod. **c**, Modulus-square of the real-space wavefunction of the $WS_2$ exciton. The plot is obtained by projecting the wavefunction onto the $WS_2$ plane, with the hole position fixed near a tungsten atom at the center of the plot. **d**, Dependence of the $g\sqrt{V}$ on the $\sqrt{N}$ (red dots), where $g$ ($\hbar\Omega/2$) is the coupling strength between the exciton and cavity and $V$ is the mode volume. A linear relationship can be obtained as predicted from the theory (red line).

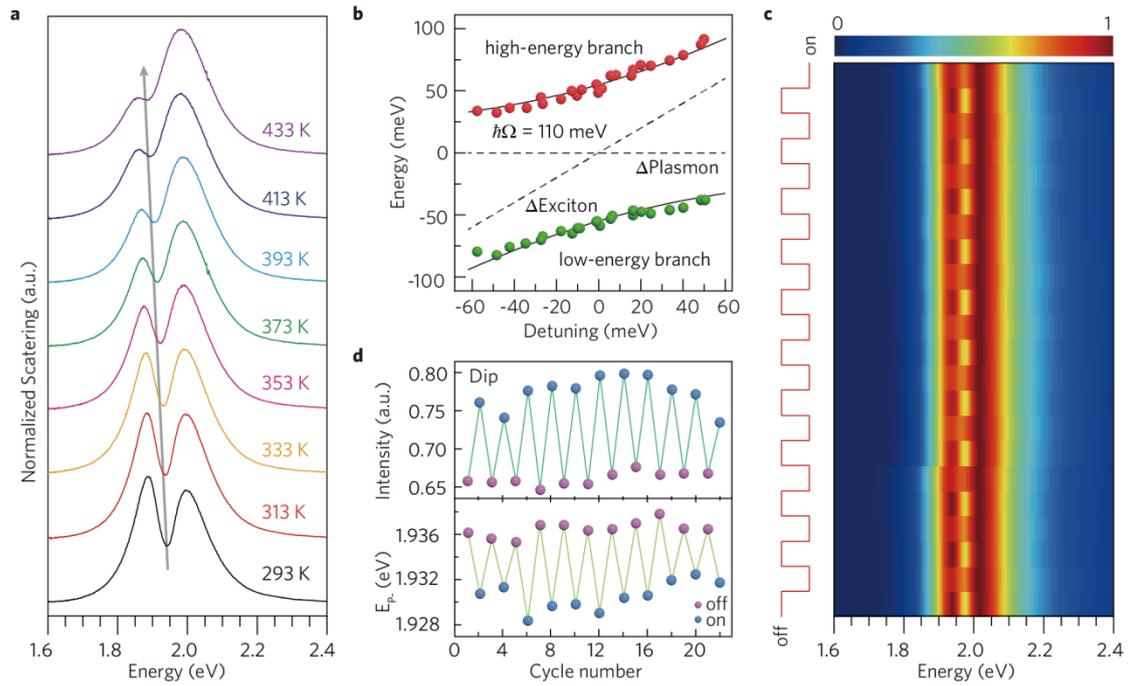

**Figure 4 | Active control of the strong coupling. a**, Temperature-dependent scattering spectra from an individual heterostructure. The temperature is scanned in 20 K steps. The grey arrow indicates the evolvement of the exciton energy with increasing temperatures. **b**, Dependence of the energy differences between the hybrid resonances and plasmon energy on the detunings between the exciton and plasmon. The red and green dots are experimental results. The black solid lines are fitting results using the coupled harmonic oscillator model. The dashed black lines indicate the detunings of the exciton transition and plasmon resonance energies with respect to the pristine room-temperature plasmon energy. **c**, Contour plot showing the evolution of the scattering spectra with alternately switching "on" and "off" the gate voltages. The color represents the scattering intensity. **d**, Cyclic performance of the dip intensity and $E_{P-}$ under an alternating −120 V external gate voltage.